\documentclass [aps,prl,twocolumn] {revtex4}
\usepackage[final]{graphics}
\usepackage{amssymb}
\usepackage{amsfonts}
\usepackage{epsfig}

\begin{document}

%\date{\today}

\hspace{15.cm} 1

\title{Effect of interfacial coupling on the magnetic ordering in ferro-antiferromagnetic bilayers}

\author{Shan-Ho Tsai$^{a,c}$, D. P. Landau$^a$, and Thomas C. Schulthess$^b$}
\affiliation{
$^a$ Center for Simulational Physics, University of Georgia, Athens, GA 30602\\
$^b$ Center for Computational Sciences and Computer Science \&
Mathematics Division, Oak Ridge National Laboratory, Oak Ridge TN 37831\\
$^c$ Enterprise Information Technology Services, University of Georgia, Athens, GA 30602}

\begin{abstract}
Monte Carlo simulations have been used to study magnetic ordering
in coupled anisotropic ferro/antiferromagnetic (FM/AFM) films
of classical Heisenberg spins. We consider films with flat interfaces
that are fully uncompensated as well as rough interfaces that are
compensated on average. For both types of interfaces above the ``N\'eel
temperature'' we observed order in the AFM with the AFM spins aligning
collinearly with the FM moments. In the case of rough interfaces
there is a transition from collinear to perpendicular alignment of
the FM and AFM spins at a lower temperature.

\end{abstract}

%\keywords{Exchange bias, anisotropy, Heisenberg spins, phase transitions}

\maketitle

\newpage

%\section{Introduction}

Ferromagnetic/antiferromagnetic (FM/AFM) bilayers have been shown to
exhibit a unidirectional shift in the hysteresis loop (exchange bias)
and a significant increase of the coercivity \cite{reviews}.
Magnetic properties of FM/AFM bilayers can also be drastically different
from those of free FM and AFM films. For example,
even though in most exchange bias systems the blocking temperature is
lower than the N\'eel temperature of the antiferromagnet, order in
the AFM has been observed above the N\'eel temperature due to the coupling
to the ferromagnet \cite{vanderzaag00}.

Earlier theoretical work \cite{koon97,schulthess98} has shown that for
compensated interfacial exchange ({\it i.e.} zero net exchange interaction
across the FM/AFM interface) the FM aligns perpendicular to the AFM easy axis.
Although similar perpendicular orientation has been observed in
numerous FM/AFM systems \cite{perp},
%neutron diffraction experiments of CoO/Fe$_3$O$_4$ multilayers \cite{ijiri98},
understanding the nature of the interfacial coupling and roughness in
FM/AFM bilayers remains a challenge.

In this paper we use Monte Carlo simulations to study the effect
of interfacial exchange interaction and roughness on transitions
from ordered to disordered states in FM/AFM bilayers. The work is
motivated by recent experiments of Ijiri et. al. \cite{ijiri02},
which show for CoO/Fe$_3$O$_4$ multilayers, that the transition to
perpendicular ordering takes place at temperatures considerably
lower than the FM and AFM ordering temperatures.

%\section{Model and Methods}
The model studied here consists of a ferromagnetic (FM) film
coupled to an underlying antiferromagnetic (AFM) film where the lattice
is coherent across the FM/AFM interface. The structure of the films
is a body-centered cubic (BCC) lattice, with linear sizes
$L_x=L_y=L\le 96$. Each film is composed of $12$ staggered (because of the
BCC structure) layers of classical unit length spins
${\bf S_r}=(S_{\bf r}^x,S_{\bf r}^y,S_{\bf r}^z)$, which interact
via the Hamiltonian
\begin{eqnarray}
{\cal H}&=&-J_F\sum_{\langle{\bf r},{\bf r'}\rangle\in {\rm FM}}{\bf S_r}\cdot {\bf S_{r'}}
-K_F\sum_{{\bf r}\in {\rm FM}}(S_{\bf r}^z)^2 \nonumber\\
&&-J_A\sum_{\langle{\bf r},{\bf r'}\rangle\in {\rm AFM}}{\bf S_r}\cdot {\bf S_{r'}}
-K_A\sum_{{\bf r}\in {\rm AFM}}(S_{\bf r}^y)^2 \nonumber \\
&&-J_I\sum_{\langle{\bf r},{\bf r'}\rangle\in {\rm FM/AFM}}{\bf S_r}\cdot {\bf S_{r'}}
\end{eqnarray}
where $\langle{\bf r},{\bf r}'\rangle$ denotes nearest-neighbor pairs of spins
coupled with exchange interactions $J_F>0$ in the FM film, $J_A<0$ in the
AFM film, and $J_I$ at the FM/AFM interface.
Spins in the AFM film have a uniaxial single-site anisotropy $K_A>0$, whose
easy axis is along the $y$ axis. The demagnetizing field on the FM film is
modeled with a hard-axis ($K_F<0$) along the $z$ direction,
which is perpendicular to the FM/AFM interfacial plane. No external magnetic
field is applied.
We use periodic boundary conditions along the $x$ and $y$ directions and
free boundary conditions along the $z$ direction.

We consider interaction and anisotropy parameters $J_F=5J>0$, $J_A=-J$,
$J_I=-J$, $K_A=J$, and $K_F=-0.5J$. We model flat interfaces as well as
``rough'' ones which are comprised of uniformly spaced steps,
with $6$, $L$, and one spin per terrace in the $x$, $y$, and $z$ directions,
respectively. The terrace sizes along the $x$ and $z$ directions are kept
fixed for the different film cross sections used.
For both types of interfaces the exchange coupling across the interface
is uniform ($J_I=-J$);
therefore, the flat interfaces are fully uncompensated while the rough ones
are compensated on average.

The FM and AFM order parameters are the uniform and staggered magnetization
per spin, respectively. These are defined as $m=|\sum_{\bf r}{\bf S_r}|/N_F$
and
$m_s=|\sum_{{\bf r}\in I}{\bf S_r}-\sum_{{\bf r}\in II}{\bf S_r}|/N_A$, where
I and II denote the two simple cubic sublattices of the BCC lattice and
$N_F$ and $N_A$ denote the number of spins on the FM and on the AFM films,
respectively. The summations here are performed over all sites on both types
of films. The magnitude of the two components of the uniform magnetization
parallel to the interfacial plane is denoted as $m_x$ and $m_y$.
Our simulations were carried out using importance sampling Monte Carlo methods
\cite{MCbook}, with Metropolis algorithm, at fixed temperature $T$.
Typically $2\times 10^5$ Monte Carlo Steps/site (MCS) were used for
computing averages after about $3 \times 10^5$ MCS were first discarded
for thermalization. Whenever not shown,
error bars in the figures are smaller than the symbol sizes.

\begin{figure}[ht]
\centering
\leavevmode
\includegraphics[clip,angle=0,width=8cm]{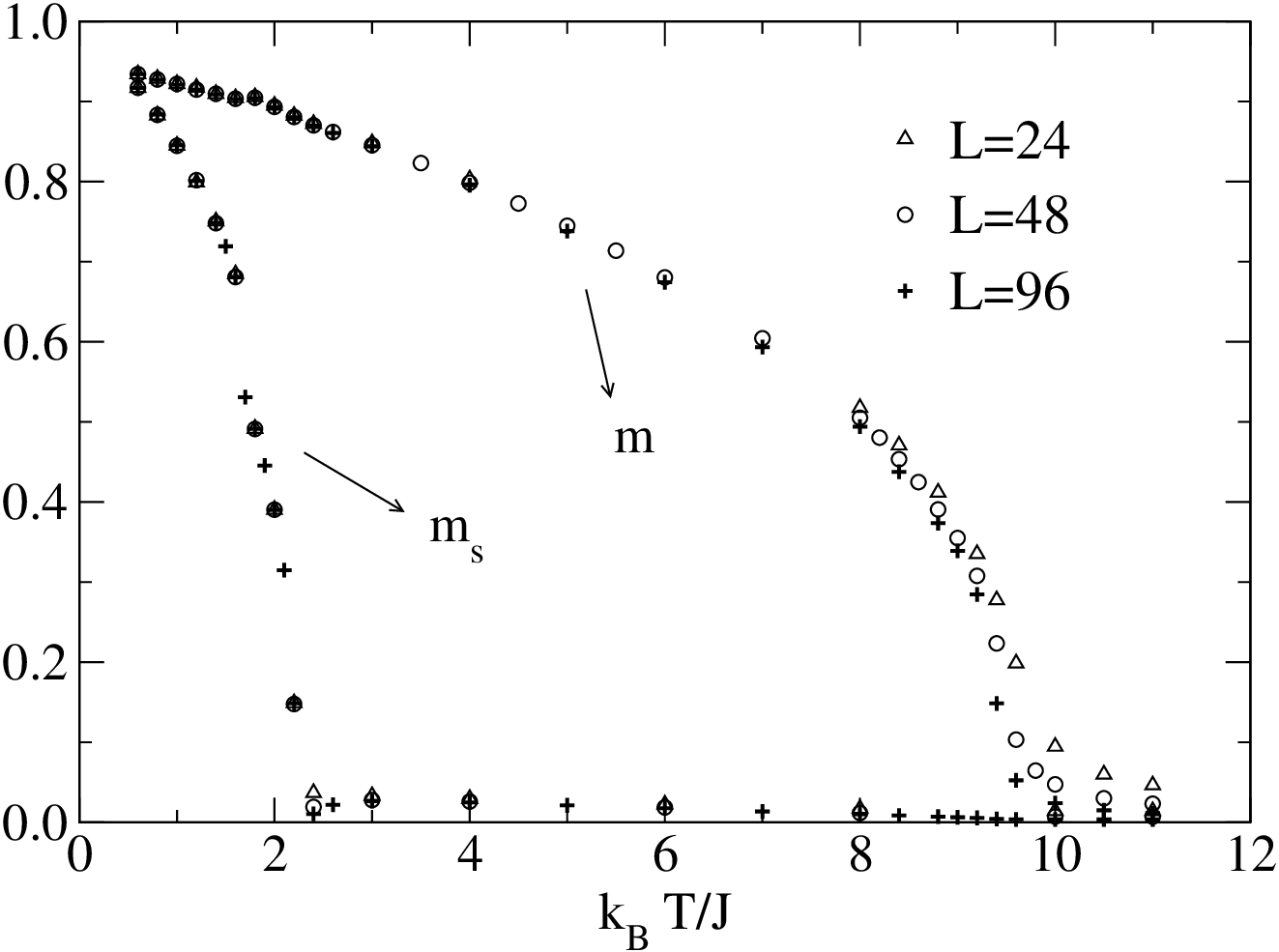}
\caption{\label{figmag}Uniform and staggered magnetization as a function of temperature for $L=24,\,48,$ and $96$, with rough interface and parameters $J_F=5J>0$, $J_A=J_I=-J$, $K_A=J$, and $K_F=-0.5J$.}
\end{figure}

%\section{Results}
Fig.\ref{figmag} shows the uniform and staggered magnetization as a function
of $T$ for systems with a rough (stepped) interface.
At low $T$, both $m$ and $m_s$ tend to unity, indicating ordered FM
and AFM films. At very high $T$, both $m$ and $m_s$ decay to zero,
indicating disordered spin configurations on both films.
Finite-size effects are illustrated using films with cross sections
$L=24,48$, and $96$, where the number of spins per terrace along the $x$
direction is constant.
These data suggest the existence of a FM transition between
ordered and disordered states at $T_c\approx 9.3J/k_B$ and a possible
N\'eel temperature at $T_N\approx 2.2J/k_B$, which coincides with the N\'eel
temperature of a free 12-atomic-layer AFM film \cite{mmm2001}.

\begin{figure}[ht]
\centering
\leavevmode
\includegraphics[clip,angle=0,width=8cm]{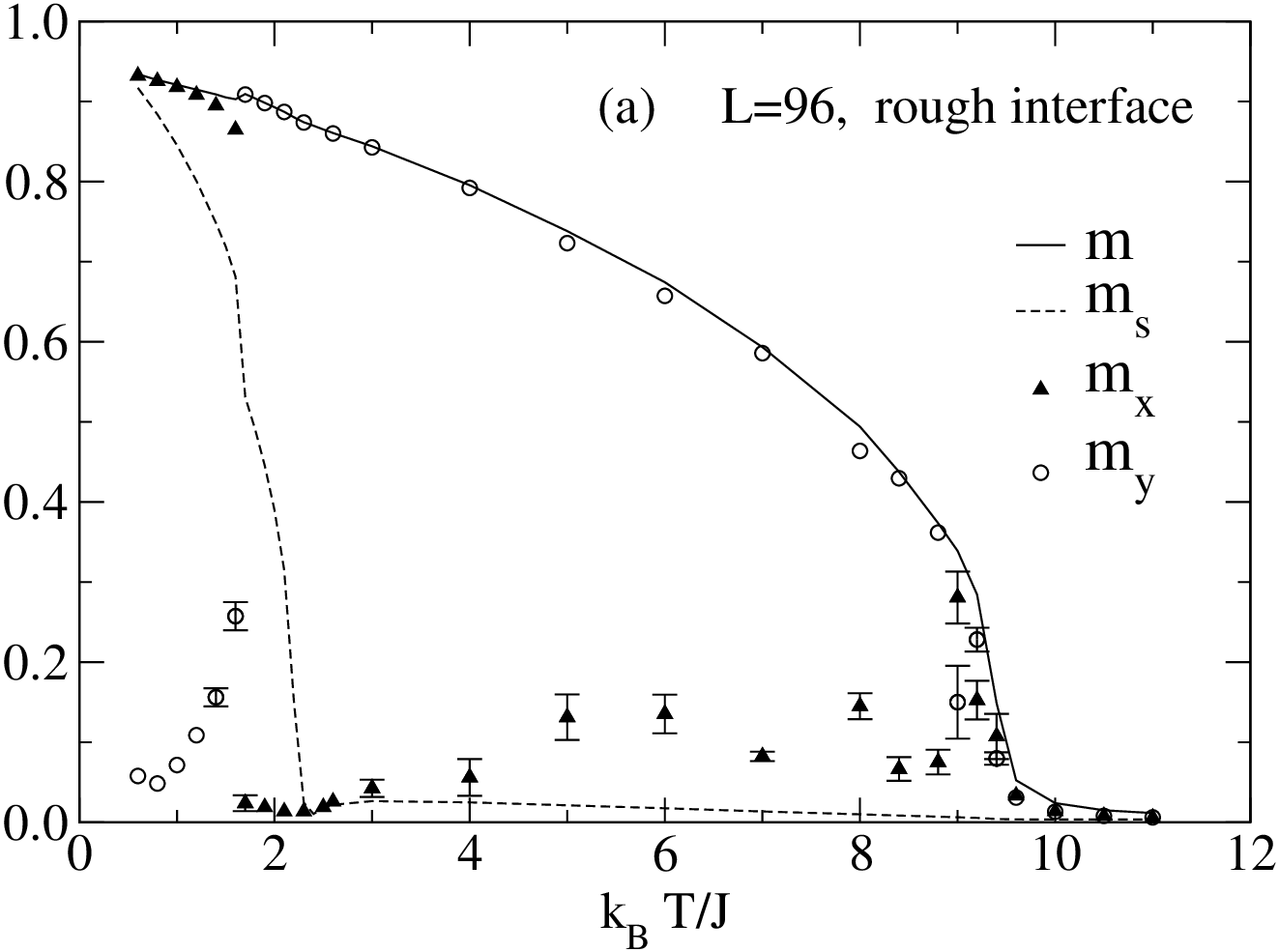}
\includegraphics[clip,angle=0,width=8cm]{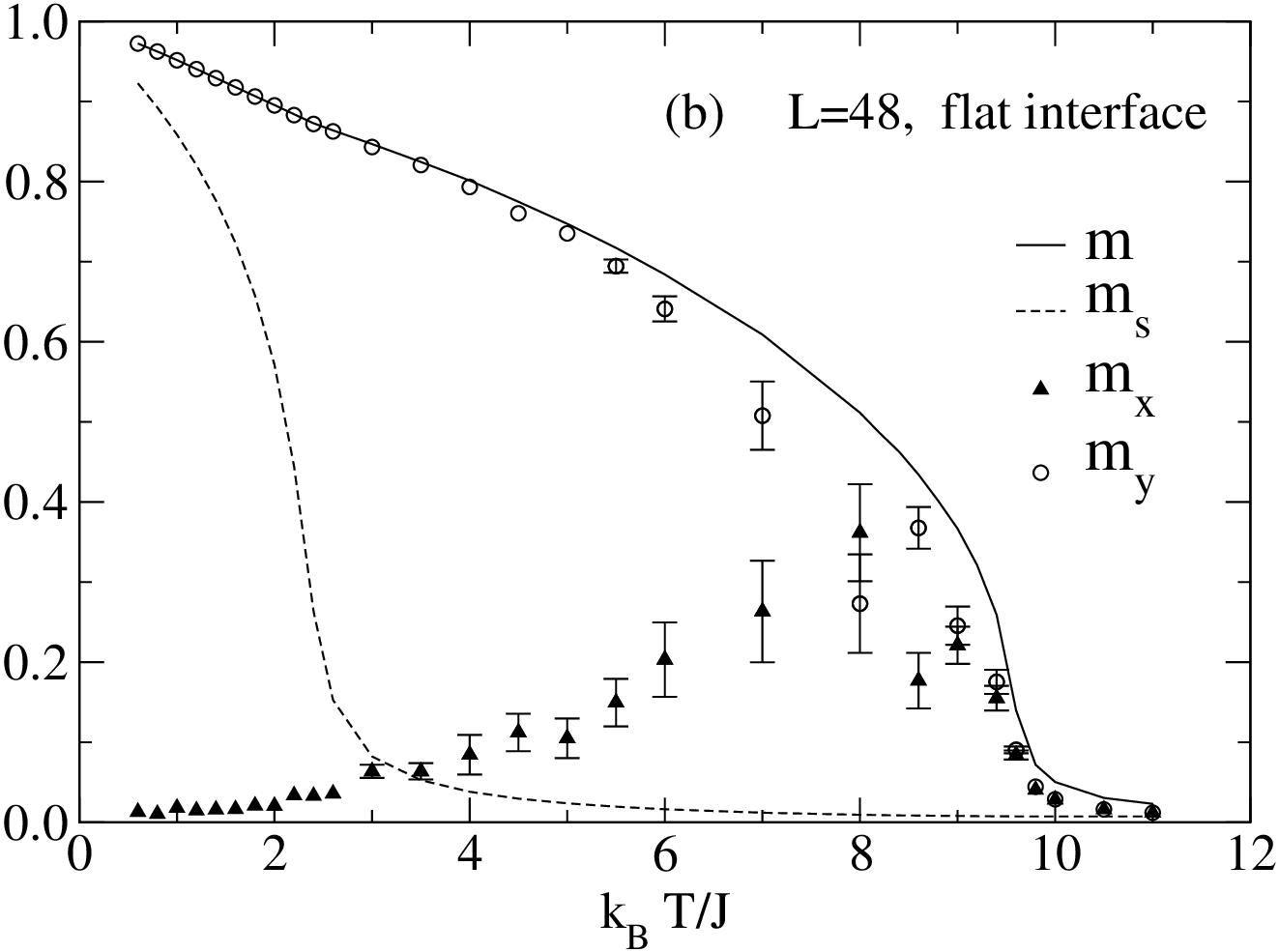}
\caption{\label{figcomp}Temperature dependence of the uniform and staggered magnetization and of the magnitude of the components of the uniform magnetization $m_x$ and $m_y$ for (a) $L=96$, with rough interface and for (b) $L=48$, with flat interface. Both cases are for $J_F=5J>0$, $J_A=J_I=-J$, $K_A=J$, and $K_F=-0.5J$.}
\end{figure}

Temperature dependences of $m_x$ and $m_y$ are shown in
Figs.\ref{figcomp}a and \ref{figcomp}b for systems with a rough and a
flat interface, respectively.
For both types of interfaces, the spins on the FM film orient
predominantly in a direction collinear with the easy axis of the
AFM at high temperatures, even above $T_N$. This is an indication
that there is still order in the AFM above $T_N$ due to the
coupling to the ferromagnet. As the temperature is lowered in the
case of the rough interface the FM spins switch to orient in a
direction that is perpendicular to the AFM easy axis, which is the
direction of the AFM spins. Our results suggest that the onset of
this perpendicular orientation is very sharp and it occurs at a
temperature below the N\'eel temperature. The $z$-components of the
uniform and staggered magnetization are very small for all $T$.
Note that in the absence of the AFM film, spins on the FM film
have global rotation symmetry in the $x$-$y$ plane, which is the
easy plane for spins on this film. The preferential orientations
of the FM spins observed here either below or above $T_N$ result
from the exchange coupling to the AFM film.

\begin{figure}[ht]
\centering
\leavevmode
\includegraphics[clip,angle=0,width=8cm]{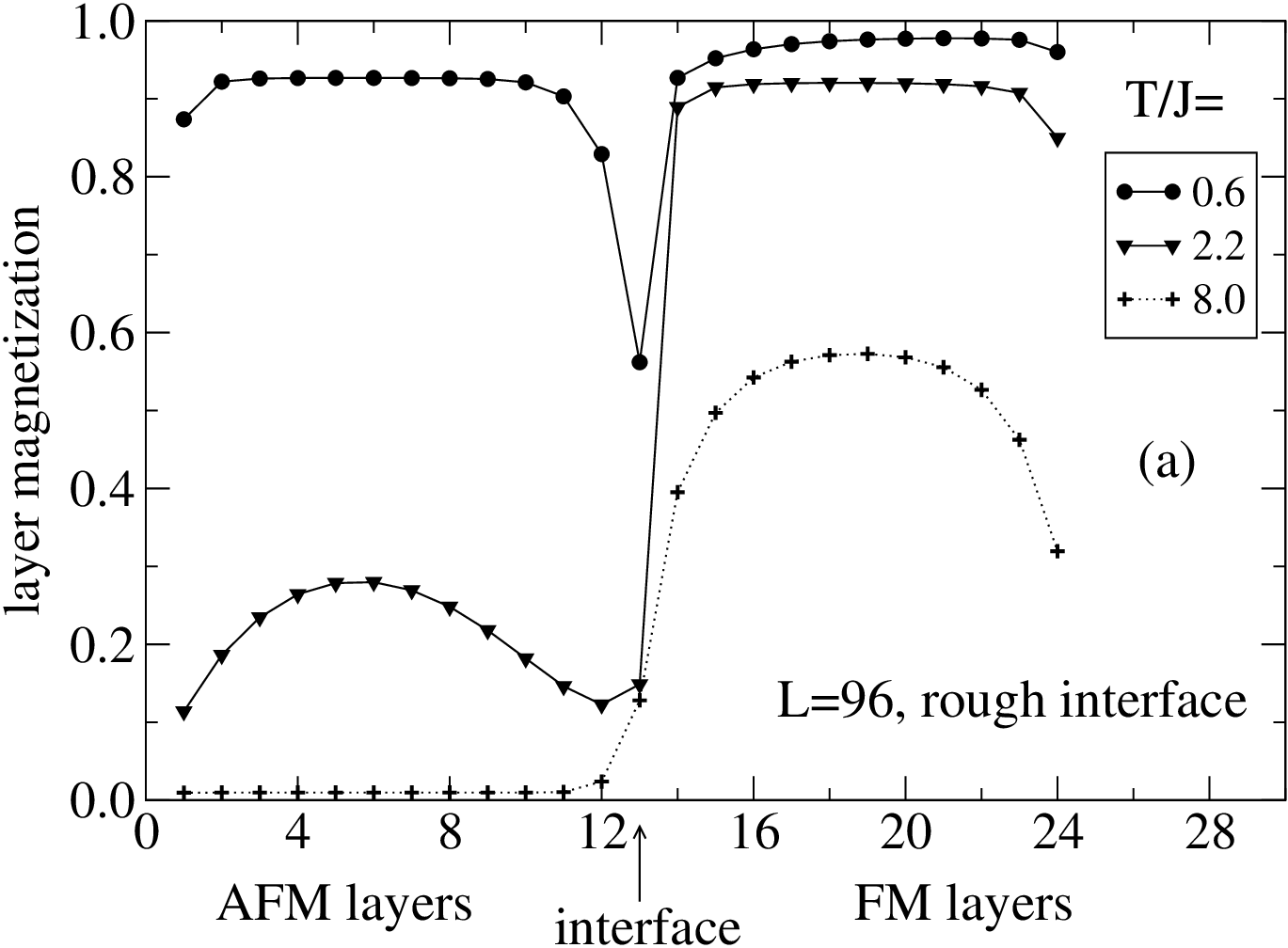}
\includegraphics[clip,angle=0,width=8cm]{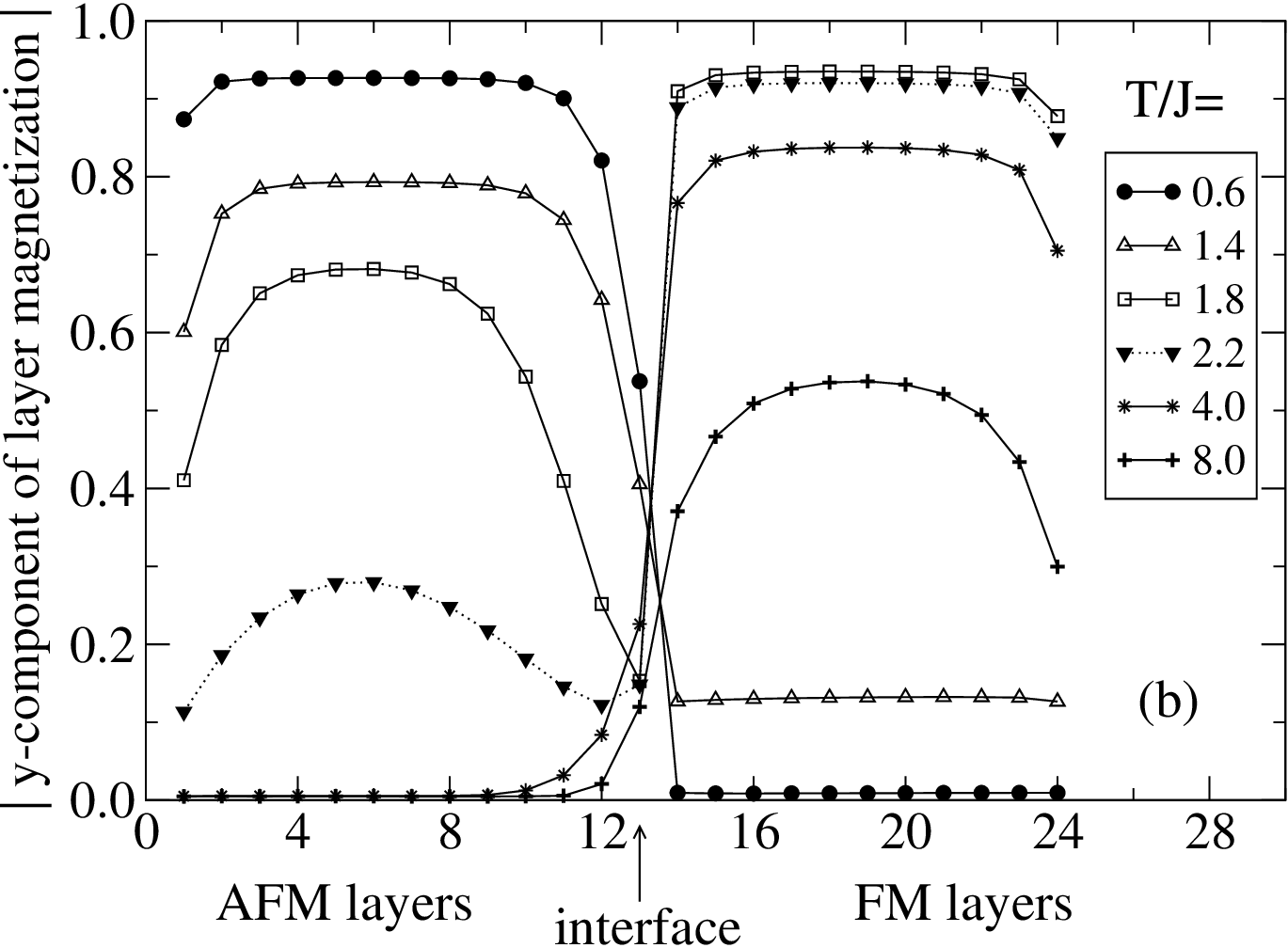}
\caption{\label{figlayrough}(a) Layer magnetization and (b) the magnitude
of its $y$-component as a function of the film layer, for rough interface
and $L=96$. Parameters used are $J_F=5J>0$, $J_A=J_I=-J$, $K_A=J$,
and $K_F=-0.5J$.}
\end{figure}

\begin{figure}[ht]
\centering
\leavevmode
\includegraphics[clip,angle=0,width=8cm]{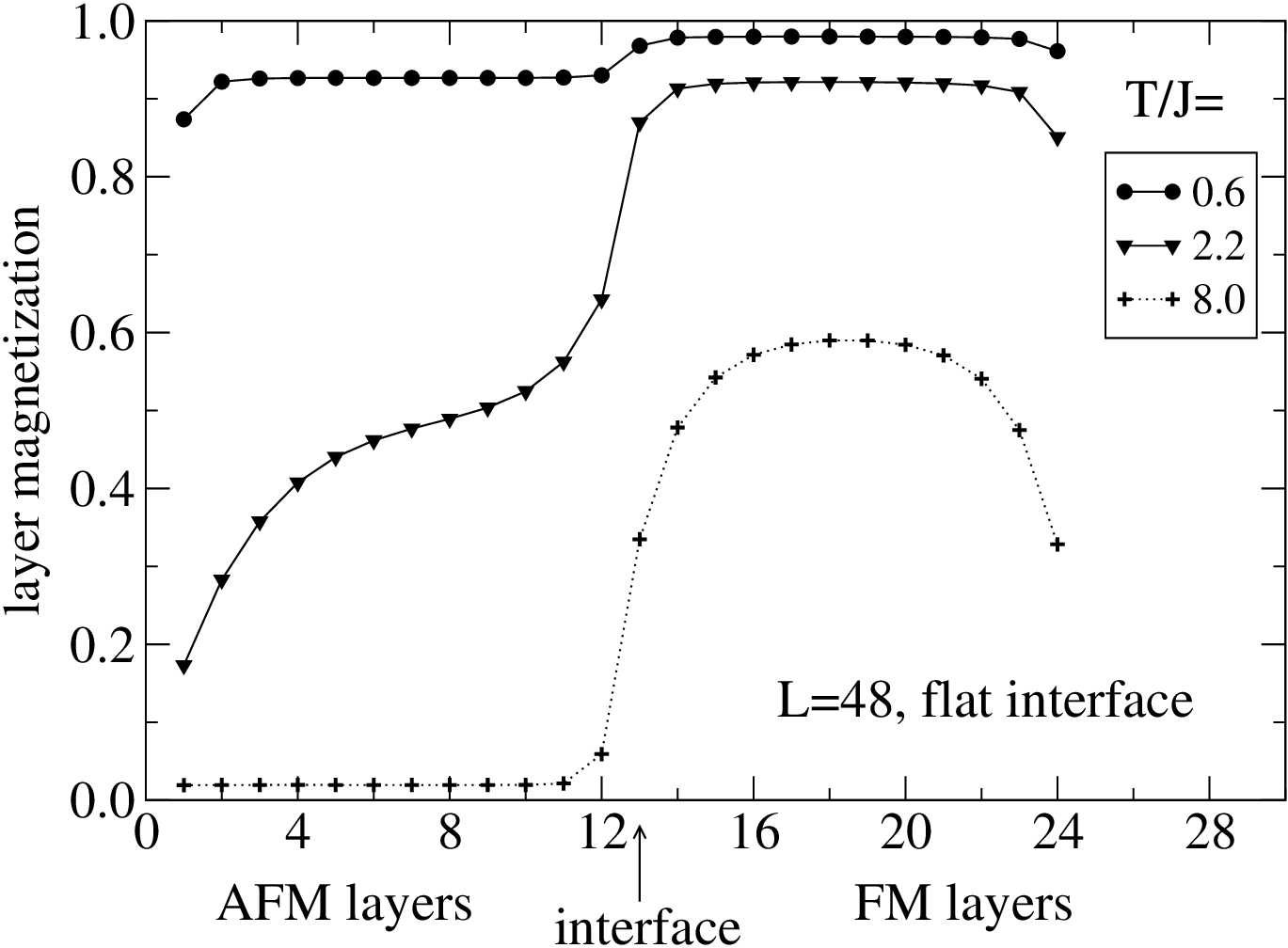}
\caption{\label{figlayflat}Layer magnetization as a function of the film
layer, for flat interface and $L=48$. Parameters used are $J_F=5J>0$,
$J_A=J_I=-J$, $K_A=J$, and $K_F=-0.5J$.}
\end{figure}

In studying thin films, it is also interesting to examine the layer
magnetization since this can vary significantly depending on the proximity
to the interface or to the free boundaries. Figs.\ref{figlayrough}a and
\ref{figlayrough}b show the layer magnetization and the magnitude of
its $y$-component, respectively, as a function of the film layer for a
system with rough interface and $L=96$.
In our notation, layers 1 to 12 comprise the AFM film,
layer 13 has uniform terraces belonging alternately to the AFM and FM films,
and layers 14 to 24 form the FM film.
For each temperature, the slightly lower magnetization
on the surface layers is a consequence of the free boundary condition along
the $z$-axis. In addition, the compensated exchange on average across the
interfacial plane reduces the layer magnetization for the layers near the
interface. The perpendicular orientation of the FM spins with the AFM spins
at low $T$ can also be seen in Fig.\ref{figlayrough}b, where {\it e.g.} for
$T=0.6J/k_B$ the magnetization on the FM layers has a very
small $y$-component. At a higher $T$, {\it e.g.} $T=1.8J/k_B<T_N$
the magnetization on the FM layers has a large $y$-component.
When the interface is flat with uncompensated exchange across it, the reduction
of the magnetization on the layers near the interface is not very large, at
low $T$ (see Fig.\ref{figlayflat}, where layers 1 to 12 comprise the AFM film
and layers 13 to 24 the FM film).

Simulations using $J_F=5J>0$, $J_A=-J$, $J_I=-2J$, $K_A=J$, and $K_F=-0.5J$
({\it i.e.} with a stronger interfacial coupling) with rough interface
suggest that the onset of the perpendicular orientation of the FM with
the AFM easy axis occurs at a lower $T$ when the interfacial coupling
increases in magnitude.
We also find that for a film with a rough interface that is compensated
on average, when the interaction and anisotropy parameters are such that
the AFM becomes disordered at a {\bf higher temperature} than the FM
({\it e.g.} with $J_F=J$, $J_A=-J$, $J_I=-J$, $K_F=-0.5J$, and $K_A=J$),
the perpendicular orientation of the FM and AFM spins is present
throughout the FM phase.

Besides the usual critical slowing down near critical points, studies of
phase transitions in thin films are further complicated by the fact that
order in the films varies layer by layer, as illustrated above. Therefore,
determining the nature of phase transitions may require a layer order
parameter. However, the goal of our current work is not to study the
thermodynamic limit of these bilayers, but rather to understand experimental
observations of finite systems.

%\section{Conclusions}
Monte Carlo simulations have been used to study a system of
coupled FM/AFM films, with flat interfaces that are fully uncompensated as
well as rough interfaces that are compensated on average. For both types of
interfaces above the N\'eel temperature we observed
order in the AFM with the AFM spins aligning collinearly with the FM moments.
As the temperature is lowered in the case of the rough interface, we have seen
a transition from collinear to perpendicular alignment of the FM and AFM spins.

%\section{Acknowledgments}
This research was partially supported by NSF grant DMR-0094422
as well as DOE-OS through BES-DMSE
and OASCR-MICS under Contract No. DE-AC05-00OR22725 with
UT-Battelle LLC. Simulations were performed on the IBM SP at NERSC.

%\newpage

\end{document}